\documentclass[prd,twocolumn,floats,aps,showpacs]{revtex4}

\usepackage{graphicx}
\usepackage{dcolumn}
\usepackage{amsmath}

\usepackage{longtable}

\begin{document}
\title{Revisiting the Vector and Axial-vector Vacuum Susceptibilities}

\author{Lei Chang}
\affiliation{Department of Physics and State Key Laboratory of
Nuclear Physics and Technology, Peking University, Beijing 100871,
China}

\author{Yu-xin Liu}
\affiliation{Department of Physics and State Key Laboratory of
Nuclear Physics and Technology, Peking University, Beijing 100871,
China} \affiliation{Center of Theoretical Nuclear Physics, National
Laboratory of Heavy Ion Accelerator, Lanzhou 730000, China}

\author{Wei-min Sun}
\affiliation{Department of Physics, Nanjing University, Nanjing
210093, China} \affiliation{Joint Center for Particle, Nuclear
Physics and Cosmology, Nanjing 210093, China}

\author{Hong-shi Zong}
\affiliation{Department of Physics, Nanjing University, Nanjing
210093, China} \affiliation{Joint Center for Particle, Nuclear
Physics and Cosmology, Nanjing 210093, China}

\date{\today}

\begin{abstract}

We re-investigate the vector and axial-vector vacuum
susceptibilities by taking advantage of the vector and
axial-vector Ward-Takahashi identities. We show analytically that,
in the chiral limit, the vector vacuum susceptibility is zero and
the axial-vector vacuum susceptibility equals three fourths of the
square of the pion decay constant. Besides, our analysis
reproduces the Weinberg sum rule.

\end{abstract}


\maketitle

The vacuum susceptibilities play an important role for
characterizing the nonperturbative aspects of quantum
chromodynamical (QCD) vacuum and in the determination of hadron
properties \cite{Ioffe,Yung,Mikhailov}. For example, the strong
and parity-violating pion-nucleon coupling depends crucially upon
the $\pi$ susceptibility \cite{Henley96}. Tensor susceptibility of
the vacuum is relevant for the determination of the tensor charge
of the nucleon \cite{He96,Belyaev97,Broniowski98,Kisslinger99,Bakulev00,Jaffe}.
Recently, in a series of
articles \cite{Zong-Vector,Zong-Axial,Zong-Tensor} the authors
have derived a closed formula for the vacuum susceptibilities,
such as the vector, axial-vector and tensor vacuum susceptibility,
using the method of differentiating the dressed quark propagator
with respect to the external field of interest. It was shown there
that the vacuum susceptibility is tightly related to the fully
dressed quark propagator and the corresponding vertex. In order to
obtain numerical values of these three kinds of vacuum
susceptibility, the authors performed corresponding calculations
in the framework of the rainbow-ladder approximation of the
Dyson-Schwinger (DS) approach. It is obvious that those
calculations depend on models, i.e., the rainbow-ladder
approximation of the DS approach. Clearly, in order to get a
reliable theoretical prediction of QCD sum rule external field
approach, one should determine the various vacuum susceptibilities
as precisely as possible. In this letter we revisit the vector and
axial-vector vacuum susceptibilities and try to obtain a
model-independent result for them in the chiral limit.

In order to make this paper self-contained, let us first recall
the definition of vacuum susceptibilities in the QCD sum rule
external field approach \cite{Ioffe,Yung,Mikhailov}. Just as was
shown in Refs. \cite{Zong-Vector,Zong-Axial,Zong-Tensor}, the QCD
vacuum susceptibility is tightly related to the linear response of
the dressed quark propagator coupled nonperturbatively to an
external current $J^{\Gamma}(y){\cal{V}}_{\Gamma}(y)\equiv
\bar{q}(y)\Gamma q(y){\cal{V}}_{\Gamma}(y)$ [$q(y)$ is the quark
field, $\Gamma$ stands for the appropriate combination of Dirac,
flavor, color matrices and ${\cal{V}}_{\Gamma}(y)$ is the variable
external field of interest]. The vacuum susceptibility
$\kappa_{\Gamma}$ in the QCD sum rule external field treatment can
be defined as
\begin{eqnarray}\label{sus_defi}
{\cal{G}}^{cc'\Gamma,NP}_{\alpha\beta}(x)&\!\!\equiv\!\! &
{\cal{G}}^{cc'\Gamma}_{\alpha\beta}(x)-
{\cal{G}}^{cc'\Gamma,PT}_{\alpha\beta}(x)\nonumber\\
&\!\!\equiv\!\! &\langle\tilde{0}|:q^{c}_{\alpha}(x)
\bar{q}^{c'}_{\beta}(0):|\tilde{0}\rangle_{J^{\Gamma}}  \nonumber \\
&\!\! \equiv \!\!
&-\!\frac{1}{12}\left(\Gamma{\cal{V}}_{\Gamma}\right)_{\alpha\beta}\delta_{cc'}
\kappa_{\Gamma}H(x)\langle\tilde{0}|\!:\!\bar{q}(0)q(0) \! : \!
|\tilde{0}\rangle \, , \nonumber \\
\end{eqnarray}
where $(cc')$, $(\alpha\beta)$ are color and spinor indices of the
quark field and the phenomenological function $H(x)$ represents
the non-locality of the two-quark condensate. Note that $H(0)=1$.
Here we follow the notations in Refs.
\cite{Zong-Vector,Zong-Axial,Zong-Tensor}. $|\tilde{0}\rangle$
denotes the exact QCD vacuum, ${\cal{G}}^{\Gamma,PT}(x)$ and
${\cal{G}}^{\Gamma,NP}(x)$ is the linear response of the dressed
quark propagator coupled perturbatively and nonperturbatively to
the external current $J^{\Gamma}(y){\cal{V}}_{\Gamma}(y)$,
respectively. In addition, the bracketting colons ``::'' are only
a notation which means that we subtract the contribution of the
perturbative term ${\cal{G}}^{\Gamma,PT}(x)$ from
${\cal{G}}^{\Gamma}(x)$. As can be seen from Eq. (\ref{sus_defi}),
in order to calculate the vacuum susceptibility $\kappa_{\Gamma}$,
one must know how to calculate ${\cal{G}}^{\Gamma,NP}(x)$ in
advance. The main task now is to determine
${\cal{G}}^{\Gamma,NP}(x)$ in a consistent way.

By differentiating the dressed quark propagator with respect to a
variable external field, the linear response of the nonperturbative
dressed quark propagator in the presence of a variable external
field can be obtained. Using this general method, the authors in
Refs. \cite{Zong-Vector,Zong-Axial} gave an expression for the
vector vacuum susceptibility $\kappa_{V}$
\begin{eqnarray}
\kappa_{V}\langle\bar{q}q\rangle & = & \int \! \! d^4z \int\!\!
\frac{d^4P}{(2\pi)^4}~e^{i P\cdot z}\frac{{\cal{V}}_{\mu
}(z){\cal{V}}_{\nu}(z)}{{\cal{V}}^{2}(z)}  \nonumber \\
&& \times \left[\Pi^{V}_{\mu\nu}(P) -\Pi^{V,PT}_{\mu\nu}(P)\right]
\, ,
\end{eqnarray}
with the vector current vacuum polarization $\Pi^{V}_{\mu\nu}(P)$
\begin{equation}\label{corre-vector}
\Pi^{V}_{\mu\nu}(P)=-Z_{2}\int^{\Lambda}_q
tr\left[\gamma_{\mu}\chi_{\nu}(q;P)\right],
\end{equation}
and that for the axial-vector one
\begin{eqnarray}
\kappa_{A}\langle\bar{q}q\rangle & =  &  \int d^4z\int
\frac{d^4P}{(2\pi)^4}~e^{i P\cdot z}\frac{{\cal{V}}_{5\mu
}(z){\cal{V}}_{5\nu }(z)}{{\cal{V}}_5^{2}(z)}
 \nonumber \\
&& \times \left[\Pi^{A}_{\mu\nu}(P)-\Pi^{A,PT}_{\mu\nu}(P)\right]
\, ,
\end{eqnarray}
with the axial-vector vacuum polarization $\Pi^{A}_{\mu\nu}(P)$
\begin{equation}\label{corre-axial}
\Pi^{A}_{\mu\nu}(P)=-Z_{2}\int^{\Lambda}_q
tr\left[\gamma_{5}\gamma_{\mu}\chi_{5\nu}(q;P)\right],
\end{equation}
where
$\int^{\Lambda}_{q}:=\int^{\Lambda}\frac{d^{4}q}{(2\pi)^{4}}$
represents mnemonically a translationally-invariant regularisation
of the integral, with $\Lambda$ the regularisation mass-scale and
the trace operation is taken over the Dirac and color indices.
Here ${\cal{V}}_{\mu }(z)$, ${\cal{V}}_{5\mu }(z)$ denote the
variable external vector and axial-vector fields.
$\chi_{\nu},\chi_{5\nu}$ denote the exact vector and axial-vector
vertex wave functions which can be defined in terms of the exact
quark propagator $S$ in the absence of the external field and the
corresponding vertex function $\Gamma$ as $S\Gamma S$. $Z_{2}$ is
the quark wave function renormalization constant.  Owing to the
Ward-Takahashi identity, the vertex renormalization constant
equals to $Z_{2}$. $\langle\bar{q}q\rangle$ denotes the chiral
quark condensate.

As is shown in Eqs. (2) and (4), in order to obtain the
nonperturbative  vector (axial-vector) vacuum susceptibility, one
should subtract $\Pi^{V,PT}_{\mu\nu}(P)$
($\Pi^{A,PT}_{\mu\nu}(P)$), which arise due to perturbative
effects. In other words, when calculating the vacuum
susceptibility, the perturbative vacuum mean values should be
subtracted. This point is very important and needs clarification.
It is well-known that the separation of perturbative and
nonperturbative contributions to vacuum mean values has some
arbitrariness. Usually, this arbitrariness is avoided by
introducing some normalization point \cite{Shifman-A,Shifman-B}.
In such a formulation, the condensates depend on the normalization
point. Other methods are also possible. For example, in the study
of the mixed quark-gluon condensate, the authors in
Ref. \cite{zong-condensate} identified the perturbative vacuum
with the Wigner vacuum, both of which are trivial in the sense
that there are no chiral symmetry breaking and confinement, in
contrast to Nambu-Goldstone vacuum (the non-trivial vacuum) which
corresponds to dynamical chiral symmetry breaking and confinement
(more detail can be found in Ref. \cite{zong-condensate}). In
Refs. \cite{Zong-Vector,Zong-Axial,Zong-Tensor}, the authors adopt
the above viewpoint to calculate the vector, axial-vector and
tensor vacuum susceptibilities in the framework of rainbow-ladder
approximation of the DS approach. For example, in the calculation
of vector vacuum susceptibility, $\Pi_{\mu\nu}^V(P)$ in Eq.~(2) is
calculated on the Nambu-Goldstone vacuum configuration, while
$\Pi_{\mu\nu}^{V,PT}(P)$ in Eq.~(2) is calculated on the Wigner
vacuum configuration. It is obvious that this calculation depends
on the rainbow-ladder approximation of the DS approach. In the
following, we shall adopt a more elegant method to study the
vector and axial- vector vacuum susceptibility (see below).

Let us first focus on the vector vacuum susceptibility
$\kappa_{V}$.  The conservation of vector current ensures that the
vector vacuum polarization is purely transverse, i.e.,
\begin{equation}\label{vector-trans}
\Pi^{V}_{\mu\nu}(P)=\left(\delta_{\mu\nu}-
\frac{P_{\mu}P_{\nu}}{P^{2}}\right)\Pi^{V}_{T}(P^{2}).
\end{equation}
Taking the constant external field limit ${\cal V}_{\mu}(z)={\cal
V}_{\mu}$  and making use of the integration formula
\begin{equation}\label{integ}
\int d^4 l~\delta^4(l)~\frac{p\cdot l k\cdot
l}{l^2}f(k,p,l)=\frac{1}{4}\int d^4 l~\delta^4(l)~p\cdot k f(k,p,l),
\end{equation}
we obtain
\begin{eqnarray} \label{sus-begin1}
\kappa_{V}\langle\bar{q}q\rangle
=\frac{3}{4}\left[\Pi^{V}_{T}(P^{2}=0) -
\Pi^{V,PT}_{T}(P^{2}=0)\right].
\end{eqnarray}
From Eq. (\ref{sus-begin1}), it can be seen that the vector vacuum
susceptibility is tightly related to the vector vacuum
polarization at zero total momentum. Contracting both sides of
Eq. (\ref{corre-vector}) with $\delta_{\mu\nu}$, one obtains
\begin{equation}\label{vector-zero}
\Pi_{T}^{V}(P^{2}=0)=-\frac{Z_{2}}{3}\int^{\Lambda}_{q} tr
\left[\gamma_{\mu}\chi_{\mu}(q;P=0) \right].
\end{equation}
In the following, we shall take advantage of the Ward-Takahashi
identity to obtain the vector vertex wave functions $\chi_{\mu}$
and therefore gives a model-independent result for the vacuum
susceptibility.

In the chiral limit, the Ward-Takahashi identity for the
vector vertex wave function is usually expressed as
\begin{equation}\label{WTI-vector}
iP_{\mu}\chi_{\mu}(q;P)=S(q_{-})-S(q_{+}),
\end{equation}
where $q_{\pm}=q\pm P/2$. Expanding the right-hand side of
Eq.~(\ref{WTI-vector}) in $P_{\mu}$ and taking the limit
$P_{\mu}\rightarrow 0$ leads to the Ward identity
\begin{equation}\label{Ward-vector}
\chi_{\mu}(q;P=0)=i\frac{\partial S(q)}{\partial q_{\mu}} \, .
\end{equation}
Substituting Eq. (\ref{Ward-vector}) into the right-hand side of
Eq. (\ref{vector-zero}) and adopting the following parametrization
of the quark propagator $S(q)$
\begin{equation}
\label{Spform} S(k) = \frac{1}{i\gamma\cdot k A(k^2) + B(k^2)} = -i
\gamma\cdot k \sigma_{V}(k^2)+ \sigma_{S}(k^2) \, ,
\end{equation}
we can obtain
\begin{eqnarray}\label{v-zero-r}
&& \Pi_{T}^{V}(P^{2}=0) \nonumber \\
&=&-\frac{Z_{2}}{3}\int_q^{\Lambda}tr \left[i\gamma_{\mu}
\frac{\partial S(q)}{\partial q_{\mu}} \right] \nonumber\\
&=&-\frac{8}{3}N_cZ_{2}\int_q^{\Lambda}\left[2\sigma_V(q^2)+q^2\frac{d \sigma_V(q^2)}{d q^2}\right].
\end{eqnarray}
Since for large $q^2$, $\sigma_V(q^2) \sim \frac{1}{q^2}$, the above integral is quadratically divergent. However, this divergence is not genuine. Note that the integrand is a total divergence, so the above
integral vanishes if a translationally invariant regularization is
adopted. Then we obtain the final result for the vector vacuum
polarization at $P^{2}=0$
\begin{equation}\label{vector-relation}
\Pi_{T}^{V}(P^{2}=0)=0.
\end{equation}
Here we note that in obtaining the above result we have taken a translationally invariant regularization. However, in actual numerical calculations of vacuum polarization in the framework of Dyson-Schwinger equations, one usually employs a cut-off to regularize the ultraviolet divergence in Eq. (\ref{corre-vector}). In this case, in order to avoid the ultraviolet divergence, one contracts the polarization tensor $\Pi_{\mu\nu}^V(P)$ with the projector ${\cal P}_{\mu\nu}^4 =\delta_{\mu\nu}-4\frac{P_\mu P_\nu}{P^2}$, which, in four dimensions, is orthogonal to $\delta_{\mu\nu}$ \cite{Roberts}. The divergent part of the integral on the right-hand-side of Eq. (\ref{corre-vector}) is proportional to $\delta_{\mu\nu}$ and hence this contraction projects out only the finite part of the integral. Following this procedure, one has
\begin{eqnarray}\label{4d-projection}
&&\Pi_T^V(P) \nonumber \\
&=&-\frac{Z_2}{3}\int_{q}^{\Lambda}tr\left[\gamma_\mu\chi_\mu(q;P)-\frac{4 \gamma\cdot P}{P^2}P\cdot\chi(q;P)\right].
\end{eqnarray}
Making use of the Ward-Takahashi identity Eq. (\ref{WTI-vector}) and taking the limit $P_\mu \rightarrow 0$ in Eq. (\ref{4d-projection}), one finds after some calculations that $\Pi_{T}^{V}(P^{2}=0)=0$, which is just Eq. (\ref{vector-relation}). Here it should be noted that physically different projection procedures (for example, projection by $\delta_{\mu\nu}$ or by $\delta_{\mu\nu}-4\frac{P_\mu P_\nu}{P^2}$) should give the same result for $\Pi_{T}^{V}(P^{2}=0)$. However, in actual numerical calculations of vacuum polarization in the framework of Dyson-Schwinger equations, a cut-off regularization is usually employed. In this case one prefers to adopt projection by $\delta_{\mu\nu}-4\frac{P_\mu P_\nu}{P^2}$ (more details can be found in \cite{Roberts} and references therein).

For the same reason the subtracted term $\Pi_{T}^{V,PT}(P^{2}=0)$
also vanishes. Thus we have reached the result that the vector
vacuum susceptibility is zero as long as the Ward identity is satisfied.
In other words, the vanishing of the vector vacuum susceptibility is a direct consequence of gauge invariance. This is closely related to the fact that photons are massless. The authors of Ref. \cite{Zong-Vector} also made use of the vector Ward identity in constructing the vertex function. They did not realize that the vector vacuum susceptibility can be expressed
with this simple formula though their numerical results implies
such a point (their obtained magnitude of vector susceptibility is
of the order $10^{-5}$).

For the case of axial-vector vacuum polarization it is a little
complicated because of the massless bound state pole existing in
the axial-vector vertex. However, the analysis is also
straightforward. The general form of the axial-vector current
vacuum polarization can be written as
\begin{equation}\label{axial-trans}
\Pi^{A}_{\mu\nu}(P)=\left(\delta_{\mu\nu} -
\frac{P_{\mu}P_{\nu}}{P^{2}}\right)\Pi^{A}_{T}(P^{2})-
\frac{P_{\mu}P_{\nu}}{P^{2}}\Pi^A_L (P^2),
\end{equation}
which contains both a transverse part and a longitudinal part.
Following the same procedure of deducing the vector vacuum
susceptibility, one obtains
\begin{eqnarray} \label{sus-begin2}
\kappa_{A}\langle\bar{q}q\rangle
&=&\frac{3}{4}\left[\Pi^{A}_{T}(P^{2}=0)-\Pi^{A,PT}_{T}(P^{2}=0)\right] \nonumber \\
&-&\frac{1}{4}\left[\Pi^{A}_{L}(P^{2}=0)-\Pi^{A,PT}_{L}(P^{2}=0)\right].
\; \;
\end{eqnarray}
From Eq. (\ref{axial-trans}) and Eq. (\ref{corre-axial}), one
obtains
\begin{eqnarray}\label{axial-component-equality}
&&\left(\delta_{\mu\nu}-
\frac{P_{\mu}P_{\nu}}{P^{2}}\right)\Pi^{A}_{T}(P^{2})-
\frac{P_{\mu}P_{\nu}}{P^{2}}\Pi^A_L (P^2) \nonumber \\
&&=-Z_{2}\int^{\Lambda}_q
tr\left[\gamma_{5}\gamma_{\mu}\chi_{5\nu}(q;P)\right].
\end{eqnarray}
In the following we shall determine the axial vector vacuum
polarization  at $P^2=0$ by using projection techniques and taking
advantage of the axial-vector Ward-Takahashi identity.

In the chiral limit, the Ward-Takahashi identity for the
axial-vector vertex can be expressed as
\begin{equation}\label{WTI-axial}
-iP_{\mu}\chi_{5\mu}(q;P)=S(q_{+})\gamma_{5}+\gamma_{5}S(q_{-}) \, ,
\end{equation}
where we also express the identity in terms of $\chi_{5\mu}$
instead of the dressed vertex $\Gamma_{5\mu}$. Since the dressed
axial-vector vertex wave function $\chi_{5\mu}$ possesses a
longitudinal massless bound-state pole, $\chi_{5\mu}$ can then be
generally expressed \cite{Maris98} as
\begin{equation}\label{axial-vertex-ge}
\chi_{5\mu}(q;P)=\gamma_{5}\chi_{\mu}^{R}(q;P)+\tilde\chi_{5\mu}(q;P)
+\frac{f_{\pi}P_{\mu}}{P^{2}}\chi_{\pi}(q;P) \,,
\end{equation}
where $\chi_{\mu}^{R}(q;P)=\gamma_{\mu}F_{R}+\gamma\cdot q q_{\mu}
G_{R}-\sigma_{\mu\nu}q_{\nu}H_{R}$ with the functions
$F_{R},G_{R},H_{R}$ being regular as $P^{2}\rightarrow 0$,
$P_{\mu}\tilde\chi_{5\mu}\sim O(P^{2})$, $f_\pi$ is the pion decay
constant in the chiral limit, and $\chi_{\pi}(q;P)$ is the
canonically normalized Bethe-Salpeter wave functions of the
massless bound state which take a general form
\begin{eqnarray}
\chi_{\pi}(q;P) & \!\! = \!\! & 2\gamma_5\left[ i E_{\pi}(q;P)
+ \gamma\cdot P F_{\pi}(q;P) \right.  \nonumber  \\
& & \left. +\gamma\! \cdot \! q q\!\cdot\! P G_{\pi}(q;P)+
\sigma_{\mu\nu} q_\mu P_\nu H_{\pi}(q;P)\right]. \;\; \;\;\;\;
\end{eqnarray}
The pion decay constant $f_{\pi}$ in the chiral limit is defined as
\begin{equation}\label{f-pion-g}
f_{\pi}P_{\mu}=Z_{2}\int_{q}^{\Lambda} tr
\left[\gamma_{5}\gamma_{\mu}\chi_{\pi}(q;P) \right].
\end{equation}
It is apparent that, as $P\rightarrow 0$, we can expand the
right-hand side of Eq. (\ref{f-pion-g}) to $O(P_{\mu})$ and obtain
the pion decay constant in the chiral limit
\begin{equation}\label{f-pion-limit}
f_{\pi}=-8N_cZ_{2}\int_{q}^{\Lambda}\left[F_{\pi}(q;0)+\frac{1}{4}q^{2}G_{\pi}(q;0)\right].
\end{equation}

Contracting both sides of Eq. (\ref{axial-component-equality})
with the projector ${\cal P}_{\mu\nu}^4$ gives
\begin{eqnarray}\label{projection-4}
&&3\Pi_T^A (P^2)+3\Pi_L^A (P^2) \nonumber \\
&=&-Z_2 \int_{q}^{\Lambda} tr \bigg\{{\cal P}_{\mu\nu}^4
\gamma_5\gamma_\mu\bigg[\gamma_5 \chi_\nu^R (q;P)
+ {\tilde \chi}_{5\nu}(q;P) \nonumber \\
&& +\frac{f_\pi P_\mu}{P^2} \chi_\pi(q;P)\bigg]\bigg\}.
\end{eqnarray}
Here we focus on the zero momentum limit $P^2=0$. After a direct
calculation, one finds that the first term in the right-hand side
of Eq. (\ref{projection-4}) vanishes. The second term also
vanishes since $P_{\mu}\tilde\chi_{5\mu}\sim O(P^{2})$. The third
term is found to be $-24f_\pi N_c Z_2
\int_{q}^{\Lambda}(F_\pi(q;0)+\frac{1}{4}q^2G_\pi(q;0))$, which
equals $3f_\pi^2$ after making use of Eq. (\ref{f-pion-limit}).
Therefore one obtains
\begin{equation}\label{transverse-longitudinal-relation}
\Pi_T^A (P^2=0)+\Pi_L^A (P^2=0)=f_\pi^2.
\end{equation}
Now contracting both sides of 
\[
\Pi^{A}_{\nu\rho}(P)=-Z_{2}\int^{\Lambda}_q
tr\left[\gamma_{5}\gamma_{\nu}\chi_{5\rho}(q;P)\right]
\]
with $\frac{P_\mu P_\rho}{P^2}$ and applying the axial-vector Ward-Takahashi identity Eq. (\ref{WTI-axial}) gives
\begin{eqnarray}\label{projection-longitudinal}
&&-\frac{P_\mu P_\nu}{P^2}\Pi_L^A(P^2) \nonumber \\
&=&-iZ_2 \frac{P_\mu}{P^2}\int_q^\Lambda tr
[\gamma_5 \gamma_\nu(S(q_+)\gamma_5+\gamma_5 S(q_-))].
\end{eqnarray}
Performing the trace operation and then expanding the right-hand-side around $P=0$ gives
\begin{eqnarray}
&&-\frac{P_\mu P_\nu}{P^2}\Pi_L^A(P^2)\nonumber  \\
&=&-4N_cZ_2\frac{P_\mu}{P^2}\int_q^\Lambda\left[P_\nu \sigma_V(q^2)+2q_\nu q\cdot P \frac{d\sigma_V(q^2)}{dq^2}+O(P^2)\right] \nonumber \\
&=&-4N_cZ_2\frac{P_\mu}{P^2}\int_q^\Lambda\left[P_\nu \sigma_V(q^2)+\frac{q^2}{2}P_\nu \frac{d\sigma_V(q^2)}{dq^2}+O(P^2)\right],
\end{eqnarray}
and therefore 
\begin{eqnarray}\label{axial-L}
\Pi_L^A(P^2=0)&=&4N_cZ_2 \int_q^\Lambda\left[\sigma_V(q^2)+\frac{q^2}{2}\frac{d\sigma_V(q^2)}{dq^2}\right] \nonumber \\ &=& 0.
\end{eqnarray}
Here we have made use of the fact that the integral in
Eq. (\ref{axial-L}) vanishes if a
translationally invariant regularization is adopted.

Combining Eq. (\ref{transverse-longitudinal-relation}) with
Eq. (\ref{axial-L}), one obtains
\begin{equation}\label{axial-component-zero-momentum}
\Pi_T^A(P^2=0)=f_\pi^2,~~~\Pi_L^A(P^2=0)=0.
\end{equation}
The subtraction term $\Pi_{T}^{A,PT}(P^{2}=0)$ and
$\Pi_{L}^{A,PT}(P^{2}=0)$ can be obtained using similar methods.
Note that the perturbative axial-vector vertex function has no
pion pole term. We obtain
\begin{equation}\label{axial-component-zero-momentum-PT}
\Pi_T^{A,PT}(P^2=0)=0,~~~\Pi_L^{A,PT}(P^2=0)=0.
\end{equation}
The axial-vector vacuum susceptibility is then found to be
\begin{eqnarray}\label{final}
\kappa_{A}\langle\bar{q}q\rangle =\frac{3}{4}f_{\pi}^{2}.
\end{eqnarray}
So far, we have derived a model-independent result for the vector and axial-vector vacuum susceptibility, which will play an important role in the related calculations of the QCD sum rule external field method. In addition, we want to stress that, as was mentioned in the forth paragraph of our paper, the separation of perturbative and nonperturbative contributions to vacuum mean values has some arbitrariness. In order to eliminate this arbitrariness, in this paper we propose a method to unambiguously separate the vector and axial-vector vacuum polarization into a perturbative and nonperturbative part. At $P^2=0$, the vector vacuum polarization tensor vanishes due to gauge invariance; and the axial-vector vacuum polarization tensor vanishes at that point as well, except in the case of dynamical chiral symmetry breaking, in which case it is related to the pion decay constant. Clearly, this is a nonperturbative effect, so there is an obvious and natural separation into a perturbative and nonperturbative part at this point.

From Eq. (\ref{vector-relation}) and
Eq. (\ref{axial-component-zero-momentum}), we find that in the
chiral limit
\begin{equation}\label{V-A}
\Pi_T^V(P^2=0)-\Pi_T^A(P^2=0)=-f_\pi^2.
\end{equation}
This result is just the Weinberg sum rule~\cite{Weinbergsumrule}.

To summarize, in this paper, based on the previous work on the
vacuum susceptibilities in
Refs. \cite{Zong-Vector,Zong-Axial,Zong-Tensor}, we re-investigate the
vector and axial-vector vacuum susceptibilities. By taking
advantage of the vector and axial-vector Ward-Takahashi
identities, we have proved that in the chiral limit the vector
vacuum susceptibility is zero and the axial-vector vacuum
susceptibility equals three fourth of the square of the pion decay
constant. From our analysis we also reproduce the Weinberg sum
rule.

\bigskip


This work was supported by the National Natural Science Foundation
of China under Contract Nos. 10425521, 10575005, 10675007,
10705002 and 10775069, the Major State Basic Research Development
Program under Contract No. G2007CB815000, the Key Grant Project of
Chinese Ministry of Education under contact No. 305001.



\begin{thebibliography}{50}

\bibitem{Ioffe} B. L. Ioffe and A. V. Smilga, Nucl. Phys. {\bf B232} (1984), 109.

\bibitem{Yung} I. I. Balitsky and A. V. Yung, Phys. Lett. {\bf B129} (1983), 328.

\bibitem{Mikhailov} S. V. Mikhailov and A. V. Radyushkin,
JETP Lett. {\bf 43} (1986), 712; Phys. Rev. {\bf D45} (1992) 1754;
A. P. Bakulev, and A. V. Radyushkin, Phys.  Lett. {\bf B271} (1991),  223.

\bibitem{Henley96} E. M. Henley, W. -Y. Hwang, and
L. S. Kisslinger, Phys. Lett. {\bf B367} (1996), 21.

\bibitem{He96} H. He, X. Ji, Phys. Rev. D {\bf 54} (1996), 6897.

\bibitem{Belyaev97} V. M. Belyaev, A. Oganesian, Phys. Lett.
{\bf B 395} (1997), 307.

\bibitem{Broniowski98} W. Broniowski, M. Polyakov, H. -C. Kim, K. Goeke,
Phys. Lett. {\bf B438} (1998), 242.

\bibitem{Kisslinger99} L. S. Kisslinger, Phys. Rev. C {\bf 59} (1999), 3377.

\bibitem{Bakulev00} A. P. Bakulev, S. V. Mikhailov, Eur. Phys. J. C
{\bf 17} (2000), 129.

\bibitem{Jaffe} R. L. Jaffe, X. Ji, Phys. Rev.
Lett. {\bf 67} (1991), 552; R. L. Jaffe, X. Ji, Nucl. Phys. {\bf
B375} (1992), 527.

\bibitem{Zong-Vector} Hong-shi Zong, Feng-yao Hou, Wei-min Sun,
Jia-lun Ping, and En-guang Zhao, Phys. Rev. {\bf C 72} (2005), 035202.

\bibitem{Zong-Axial} Hong-shi Zong, Yuan-mei Shi, Wei-min Sun, and Jia-lun
Ping, Phys. Rev. {\bf C 73} (2006), 035206.

\bibitem{Zong-Tensor} Yuan-mei Shi, Kong-ping Wu, Wei-min Sun,
Hong-shi Zong and Jia-lun Ping, Phys. Lett. {\bf B 639} (2006), 248.

\bibitem{Shifman-A} V. A. Novikov, M. A. Shifman, A. I. Vainstein,
V. I. Zakharov, Nucl. Phys. {\bf B 249} (1985), 445.

\bibitem{Shifman-B} M. A. Shifman, Prog. Theor. Phys. Suppl. {\bf 131} (1998), 1.

\bibitem{zong-condensate} Hong-shi Zong, Jia-lun Ping, Hong-tin Yang,
Xiao-fu L\"{u}, Fan Wang, Phys. Rev. {\bf D 67} (2003), 074004.

\bibitem{Roberts} Conrad J Burden, Justin Praschifka,and Craig D. Roberts,
Phys. Rev. {\bf D 46} (1992), 2695.

\bibitem{Maris98} Pieter Maris, Craig D. Roberts and Peter C. Tandy,
Phys. Lett. {\bf B 420} (1998), 267.

\bibitem{Weinbergsumrule} S. Weinberg, Phys. Rev. Lett. {\bf 18} (1967), 507.

\end{thebibliography}
\end{document}